\documentclass[twocolumn,aps,superscriptaddress,showpacs,floatfix,prd,noshowpacs]{revtex4-2}
\usepackage{mathrsfs}
\usepackage{amssymb}
\usepackage{amsmath}
\usepackage{graphicx}
\usepackage[normalem]{ulem}
\usepackage[dvips]{color}
\usepackage{bm}
\usepackage{longtable}
\usepackage{slashed}
\usepackage{enumitem}
\usepackage{empheq}

\usepackage{titlesec}
\renewcommand{\thesection}{\Roman{section}}
\titleformat{\section}{\small\bfseries\centering}{\thesection.}{0.5em}{}

\usepackage{utopia}
\usepackage[T1]{fontenc}
\usepackage[utopia]{mathdesign}

\usepackage[breaklinks=true]{hyperref}
\hypersetup{
  colorlinks=true,
  citecolor=magenta,
  linkcolor=black,
  urlcolor=magenta
}

\renewcommand\sout{\bgroup \color{red} \ULdepth=-.5ex \ULset}

\renewcommand{\rm}[1]{\textrm{#1}}
\renewcommand{\d}{\mathrm{d}}

\usepackage{tikz,xcolor,hyperref}

\definecolor{lime}{HTML}{A6CE39}
\DeclareRobustCommand{\orcidicon}{
	\begin{tikzpicture}
	\draw[lime, fill=lime] (0,0) 
	circle [radius=0.16] 
	node[white] {{\fontfamily{qag}\selectfont \tiny ID}};
	\draw[white, fill=white] (-0.0625,0.095) 
	circle [radius=0.007];
	\end{tikzpicture}
	\hspace{-2mm}
}
\foreach \x in {A, ..., Z}{%
	\expandafter\xdef\csname orcid\x\endcsname{\noexpand\href{https://orcid.org/\csname orcidauthor\x\endcsname}{\noexpand\orcidicon}}
}

\begin{document}

\newcommand{\x}{\mathrm{X}}
\newcommand{\hr}{\widehat{r}}
\newcommand{\hP}{\widehat{P}}
\newcommand{\heps}{\widehat{\varepsilon}}

\title{Is the Trace Anomaly at its Minimum Value at Neutron Star Centers?}

\author{Bao-Jun Cai\orcidA{}}\email{bjcai@fudan.edu.cn}
\affiliation{Key Laboratory of Nuclear Physics and Ion-beam Application (MOE), Institute of Modern Physics, Fudan University, Shanghai 200433, China} 
\affiliation{Shanghai Research Center for Theoretical Nuclear Physics, NSFC and Fudan University, Shanghai 200438, China}
\author{Bao-An Li\orcidB{}}\email{Bao-An.Li$@$etamu.edu}
\affiliation{Department of Physics and Astronomy, East Texas A\&M University, Commerce, TX 75429-3011, USA}
\author{Yu-Gang Ma\orcidC{}}\email{mayugang$@$fudan.edu.cn}
\affiliation{Key Laboratory of Nuclear Physics and Ion-beam Application (MOE), Institute of Modern Physics, Fudan University, Shanghai 200433, China} 
\affiliation{Shanghai Research Center for Theoretical Nuclear Physics, NSFC and Fudan University, Shanghai 200438, China}

\date{\today}

\begin{abstract}

While the equation of state (EOS) $P(\varepsilon)$ of neutron star (NS) matter has been extensively studied, the EOS-parameter $\phi = P/\varepsilon$ or equivalently the dimensionless trace anomaly $\Delta = 1/3 - \phi$, which quantifies the balance between pressure $P$ and energy density $\varepsilon$, remains far less explored, especially in NS cores. 
Its bounds and density profile carry crucial information about the nature of superdense matter. 
Physically, the EOS-parameter $\phi$ represents the mean stiffness of matter accumulated from the stellar surface up to a given density. Based on the intrinsic structure of the Tolman--Oppenheimer--Volkoff equations, we show that $\phi$ decreases monotonically outward from the NS center, independent of any specific input NS EOS model. Furthermore, observational evidence of a peak in the speed-of-sound squared (SSS) density-profile near the center effectively rules out a valley and a subsequent peak in the radial profile of $\phi$ at similar densities, reinforcing its monotonic decrease. These model-independent relations impose strong constraints on the near-center behavior of the EOS-parameter $\phi$, particularly demonstrating that the mean stiffness (or equivalently $\Delta$) reaches a local maximum (minimum) at the center.
\end{abstract}

\maketitle

\section{Introduction}

Neutron stars (NSs) harbor the densest visible matter in our Universe, providing a unique laboratory to probe strongly interacting matter under strong-field gravity at supranuclear densities\,\cite{Walecka1974,Collins1975,Chin1976,Freedman1977,Akmal1998,LP01,Alford2008,LCK08,Baym18, Isa18,LCXZ21}. The cold matter equation of state (EOS)\,\cite{Landau1987},  $P = P(\varepsilon)$, which relates pressure $P$ to energy density $\varepsilon$,  governs NS internal structures and global properties\,\cite{Shapiro1983}, determining key observables such as the mass-radius (M-R) relation, tidal deformability and the maximum mass of NSs. The EOS is also crucial for interpreting heavy-ion collision data\,\cite{LCK08,Soren2023}, exploring nuclear structure problems\,\cite{Bend03,ZC16,Ding24,Yue24,Qin25}, and understanding extreme astrophysical events or processes such as supernovae and NS mergers\,\cite{Oertel2017,Chat24}. While great progress has been made in narrowing down the constraining band on $P(\varepsilon)$\,\cite{Tews18,Baym19,McL19,Zhao20,Tan22,Tan22-a,Alt22,Dri22,Huang22,Ecker22,Ecker23,Pro23,Som23,
Ann18,Ann23,Ess21,Brandes23,Brandes23-a,Tak23,Pang23,Fan23,Bed15,Cai17,Olii23,ZLi25,Komo23,Marc23,Xie20-a,Xie21,Xie23,Jim24,LBA24,ZNB25,Grun25-a,LBA25,Ecker25,Huang25,Patra25,Ng25,Bisw25,Reed25,Pass25,Wout25,Baus25,Tang25,Cui25} (see reviews\,\cite{Wat16,Dri21,Lovato22,LiA25}) utilizing observational data especially since GW 170817\,\cite{Abbott2017,Abbott2018,Abbott2020-a,Riley19,Miller19,Fon21,Riley21,Miller21,Salmi22,Choud24,Reardon24,Shirke25},  surprisingly much less is known about the behavior of the EOS-parameter (dimensionless ratio)
\begin{equation}
\phi = {P}/{\varepsilon}.
\end{equation}
With respect to the normally extracted constraining band on $P(\varepsilon)$, the EOS-parameter $\phi$ encodes complementary and physically rich information about the superdense matter in NSs. Indeed, there are interesting questions about $\phi$ to be studied. For example, does $\phi$ have bounds, or is it monotonically increasing or decreasing? Addressing such questions can deepen our understanding about the nature of superdense matter.

As a reference, in the exploration of dark energy\,\cite{Carr01,Peeb03,Cope06}, the EOS can be conveniently characterized by the EOS-parameter $w = P/\varepsilon$,  which compactly encapsulates the dynamical properties of dark energy: $w=-1$ corresponds to a cosmological constant $\Lambda$ with constant energy density $\varepsilon_{\Lambda}$ and negative pressure $P_{\Lambda}=-\varepsilon_\Lambda$, whereas any deviation from this value signals a dynamical dark energy component (such as the quintessence or chameleon models) or a modification of general relativity. The determination of $w$ thus provides a direct probe of the stiffness and possible evolution of dark energy. Fundamentally, knowing better the $w$ remains central to understanding the physical origin of cosmic acceleration\,\cite{Koma11,Aubo15,Agha20}.
While the $w$ in dark energy studies is a global quantity, the EOS-parameter $\phi$ in NSs depends on the local energy density, or equivalently on the radial coordinate, with the underlying physics expected to be richer and more intricate.

The EOS-parameter $\phi$ can be interpreted as the average sound-speed squared (SSS) from the NS surface to a given radial point\,\cite{Saes24,Marc24}:
\begin{equation}
\phi(\varepsilon)=\frac{1}{\varepsilon} \int_0^{\varepsilon} s^2(\varepsilon') \d\varepsilon'=
\langle s^2(\varepsilon)\rangle
, \quad s^2 = \frac{\d P}{\d\varepsilon}=\frac{\d\hP}{\d\heps},
\end{equation}
where $\hP$ and $\heps$ represent the pressure and energy density scaled by the central energy density (see Section \ref{SEC_2}).
In this sense, $\phi$ captures how the stiffness of matter accumulates from the surface to the inner position under consideration, effectively providing a global measure of the internal sound-speed profile. While $P(\varepsilon)$ constrains local properties, $\phi$ reveals the integrated response of matter to compression, serving as a powerful probe of dense matter properties.
By definition, $\phi$ depends on the radial distance from the center, even though it represents an integrated quantity.

Studies of $\phi$ are fully equivalent to analyzing the dimensionless trace anomaly\,\cite{Fuji22},
\begin{equation}
\Delta =1/3- \phi,
\end{equation}
which quantifies deviations from the conformal limit of superdense matter. In perturbative QCD (pQCD), $\Delta$ is predicted to approach zero from above at asymptotically high densities, reflecting the approximate restoration of conformal symmetry of quark matter\,\cite{Bjorken83,Kurkela10,Gorda21PRL,Gorda23PRL,Gorda23,Komo22}. Understanding how $\Delta$ (or equivalently $\phi$) behaves inside NS cores thus connects nuclear physics, astrophysics, and fundamental QCD in a single quantity\,\cite{Braun2022,Brandt2023,JHChen24,Fuku2024}.

In order to study the EOS-parameter $\phi$ in a model-independent manner, we employ the IPAD-TOV approach (intrinsic and perturbative analyses of dimensionless Tolman--Oppenheimer--Volkoff (TOV) equations)\,\cite{CLZ23-a,CLZ23-b,CL24-a,CL24-b,CL25-b,CL25-a}. The TOV equations\,\cite{TOV39-1,TOV39-2,Misner1973} govern the radial evolution of pressure and energy density in static, spherically symmetric NSs under hydrodynamic equilibrium in general relativity, and form the fundamental basis for NS studies.
By introducing reduced pressure and energy density normalized to the central energy density, the TOV equations can be recast into dimensionless forms, exposing more clearly their intrinsic structures suitable for input EOS-independent analyses. Unlike conventional input EOS-based approaches for solving numerically the original TOV equations, the IPAD-TOV framework expresses the reduced variables as polynomials of the reduced radius. Importantly, this enables the TOV equations to enforce intrinsic relations among the polynomial coefficients, thereby revealing the central EOS underlying a given NS mass and/or radius without using any input EOS model; see Ref.\,\cite{CL25-a} for a comprehensive review. Indeed, this method has yielded new insights into NS core properties directly from observables, such as novel scalings of NS radius and mass and the revised causality boundary of the M-R relation\,\cite{CLZ23-a,CLZ23-b,CL24-a,CL24-b,CL25-b}. Here it is employed to study the density (radial) profiles of the EOS-parameter $\phi$ and trace anomaly $\Delta$ in NS cores.

Within the region where IPAD-TOV is valid, we show in this work that the EOS-parameter $\phi$ decreases monotonically outward from the NS center, independent of any specific input EOS model for NS matter. Furthermore, a possible peak in $s^2$ near the center effectively rules out a valley or subsequent peak in $\phi$ at similar densities/positions, which provides additional support for its monotonic decrease.  Equivalently, this corresponds to a monotonic increase of $\Delta$, a simple trend that may offer useful insights into the properties of superdense matter in NSs.

It should be noted that the IPAD-TOV approach is perturbatively effective and most reliable near the NS center. Its applicability to outer regions, where the EOS-parameter $\phi$ might develop peaks or valleys, is limited, and such features require full EOS-dependent TOV solutions.

The remainder of this paper is organized as follows. Section \ref{SEC_2} briefly reviews the IPAD-TOV framework. Section \ref{SEC_3} demonstrates the monotonic outward decrease of $\phi$ near NS centers. Section \ref{SEC_4} shows that a local minimum in $\phi$ is incompatible with a local peak in the SSS $s^2$ density profile in NSs at the TOV configuration (located at the peaks of the M-R curves).  Section \ref{SEC_5} generalizes the discussion to NSs below the TOV configuration.  Section \ref{SEC_6} discusses pQCD constraints at extremely high densities and their potential implications for $\phi$ and $\Delta$ in NSs. The complementary mean stiffness of NS core matter is introduced in Section \ref{SEC_7} to complete the relevant analyses of the EOS-parameter and trace anomaly. Section \ref{SEC_8} presents our conclusions.

\section{Brief Review of the IPAD-TOV Framework}\label{SEC_2}

In this section, we briefly review the IPAD-TOV approach\,\cite{CLZ23-a,CLZ23-b,CL24-a,CL24-b,CL25-b,CL25-a}, focusing on the features most relevant for our study of $\phi$. In units $c=G=1$, the dimensionless TOV equations are\,\cite{CLZ23-a}
\begin{equation}
\frac{\d\widehat{P}}{\d\widehat{r}} = - \frac{\widehat{\varepsilon}\widehat{M}}{\widehat{r}^2} 
\frac{(1+\widehat{P}/\widehat{\varepsilon}) (1 + \widehat{r}^3 \widehat{P}/\widehat{M})}{1 - 2\widehat{M}/\widehat{r}}, \quad
\frac{\d\widehat{M}}{\d\widehat{r}} = \widehat{r}^2 \widehat{\varepsilon},
\end{equation}
where the reduced quantities are defined as $\widehat{P} = P/\varepsilon_{\rm c}$, $\widehat{\varepsilon} = \varepsilon/\varepsilon_{\rm c}$, $\widehat{r} = r/Q$ and $\widehat{M} = M/Q$, with $\varepsilon_{\rm c}$ the central energy density and $Q$ the characteristic length scale\,\cite{CL25-a}
\begin{equation}
Q \equiv \frac{1}{\sqrt{4\pi \varepsilon_{\rm c}}} 
\approx10\cdot\left(\frac{\varepsilon_{\rm{c}}\text{ in }\rm{MeV}/\rm{fm}^3}{600}\right)^{-1/2}\,\rm{km},
\end{equation}
therefore $Q$ is on the order of $\mathcal{O}(10\,\rm{km})$.
The reduced NS radius is defined by the condition $\widehat{P}(\widehat{R}) = 0$, and the corresponding NS mass is
\begin{equation}
\widehat{M}_{\rm{NS}} \equiv \widehat{M}(\widehat{R}) = \int_0^{\widehat{R}} \d\widehat{r}\, \widehat{r}^2 \widehat{\varepsilon},\text{ or, }M_{\rm{NS}}\equiv M(R).
\end{equation}

Typical NSs have radii of about 12-14\,km, meaning that 
\begin{equation}\label{RO1}
\widehat{R}=R/Q\approx\mathcal{O}(1).
\end{equation}
Similarly, we can obtain $\widehat{M}_{\rm{NS}}=M_{\rm{NS}}/Q\lesssim0.5$ considering the empirical facts  $M_{\rm{NS}}/M_{\odot}\lesssim2.5$\,\cite{Rezz18,Marg17,Ruiz18,Shib18,Zhou19,Jiang20,Shao20,Raai21,Tang21,Muso24,Koehn24,Snep24} and $\varepsilon_{\rm{c}}\lesssim1.2\,\rm{GeV}/\rm{fm}^3$\,\cite{Latt05,Latt11} holding for realistic NSs.  The smallness of $\widehat{M}_{\rm{NS}}\lesssim0.5$ and $\widehat{R} \approx \mathcal{O}(1)$ is in fact the manifestation of the compactness $\widehat{M}_{\rm{NS}}/\widehat{R}$ of NSs\,\cite{Ann22,Lind84,Latt90,Link99}.
A point located very close to the center of the NS can be roughly described by:
\begin{equation}
\text{NS core near the center: }\hr \lesssim \left(\tfrac{1}{4}-\tfrac{1}{3}\right)\widehat{R}\lesssim0.4.
\end{equation}
These dimensionless estimates provide useful intuition for the order-of-magnitude of characteristic quantities in NSs. In the following, when referring to the NS cores, we generally mean regions that satisfy this condition.

Near the NS center, there are two relevant small quantities: the reduced radius $\widehat{r}$ (or equivalently $\mu \equiv \widehat{\varepsilon} - \widehat{\varepsilon}_{\rm c} = \widehat{\varepsilon} - 1$) and the central EOS-parameter $\x \equiv\phi_{\rm{c}}\equiv \widehat{P}_{\rm c} <1$. These small quantities allow a general double-element expansion of any NS quantity $\mathcal{U}$\,\cite{CL25-a}:
\begin{equation}
{\mathcal{U}}/{\mathcal{U}_{\rm c}} \approx 1 + \sum_{i+j \ge 1} u_{ij} \x^i \widehat{r}^j,
\end{equation}
where $\mathcal{U}_{\rm c}$ is the central value and the coefficients $\{u_{ij}\}$ are determined from the TOV equations and in general the input EOS. While higher-order (near the surface) coefficients depend on the input EOS, the lower-order (near the center) coefficients are input EOS-independent\,\cite{CLZ23-b}, and it is this feature that allows us to gain robust insights into supderdense matter without using any input NS EOS.  In the limit $\widehat{r} \to 0$, the expansion becomes exact, providing a reliable and essentially model-independent framework for studying the EOS in the NS core.

Applied to the reduced pressure, the pertubative expansion reads\,\cite{CLZ23-a}
\begin{equation}
\widehat{P}(\widehat{r}) \approx \x + b_2 \widehat{r}^2 + b_4 \widehat{r}^4 +b_6\hr^6+ \cdots,
\end{equation}
with\,\cite{CLZ23-a}
\begin{equation}
b_2 = -\frac{1}{6} \left(1 + 3 \x^2+ 4 \x \right), \quad
b_4 = -\frac{1}{2} b_2 \left( \x + \frac{4+9\x}{15 s_{\rm c}^2} \right),
\end{equation}
and\,\cite{CL25-a}
\begin{align}
b_6=&-\frac{1}{216}\left(1+9\x^2\right)\left(1+3\x^2+4\x\right)-\frac{a_2^2}{30}\notag\\
&+\left(\frac{2}{15}\x^2+\frac{1}{45}\x-\frac{1}{54}\right)a_2-\left(\frac{5+12\x}{63}\right)a_4,\label{ee-b6}
\end{align}
where $s_{\rm c}^2 = [\d\widehat{P}/\d\widehat{\varepsilon}]_{\rm c}=b_2/a_2$ is the central SSS. 
In Eq.\,(\ref{ee-b6}), $a_2$ and $a_4$ are the expanding coefficients of the energy density:
\begin{equation}
\widehat{\varepsilon}(\hr) \approx 1 + a_2 \widehat{r}^2 + a_4 \widehat{r}^4 + a_6\hr^6+\cdots.
\end{equation}
By symmetry, only even powers of $\widehat{r}$ appear in the perturbative expansions of $\hP$ and $\heps$\,\cite{CLZ23-a,CLZ23-b}.
The coefficient $b_2$ is definitely negative and $b_4$ is positive-definite; consequently $a_2=b_2/s_{\rm{c}}^2<0$ and $a_4$ is possibly positive or negative.
Considering that $|a_2|\sim\mathcal{O}(1)$ and $|a_4|\sim\mathcal{O}(1)$ one can see that the magnitude of $b_6$ is smaller than that of $b_2$ or $b_4$; this indicates the convergence of $\hP$ on the expansion over $\hr$.
All coefficients are naturally of order unity due to the dimensionless expansion.

Truncating at $\mathcal{O}(\widehat{r}^2)$ yields scalings for $\widehat{R}^2$\,\cite{CLZ23-a}:
\begin{equation}
\widehat{R}^2 \sim \frac{\x}{1  + 3\x^2+ 4 \x }, 
\end{equation}
and consequently the NS radius scaling\,\cite{CLZ23-a}:
\begin{equation}
R \sim
\frac{\widehat{R}}{\sqrt{\varepsilon_{\rm{c}}}}
\sim \frac{1}{\sqrt{\varepsilon_{\rm c}}} \left(\frac{\x}{1  + 3\x^2+ 4 \x }\right)^{1/2},
\end{equation}
as well as the mass scaling\,\cite{CLZ23-a}:
\begin{equation}
M_{\rm{NS}} \sim 
\frac{\widehat{M}_{\rm{NS}}}{\sqrt{\varepsilon_{\rm{c}}}}\sim
\varepsilon_{\rm c} R^3 \sim \frac{1}{\sqrt{\varepsilon_{\rm c}}} \left(\frac{\x}{1 + 3\x^2+ 4 \x }\right)^{3/2}.
\end{equation}

These relations directly connect global NS properties to the central EOS-parameter $\x$, without relying on the detailed composition or the specific input EOS model, as reflected by the absence of higher-order coefficients such as $a_4$\,\cite{CLZ23-b,CL24-a}. In practice, the mass scaling is more robust than the radius, since the latter is more affected by the uncertain low-density EOS in the NS crust, which contributes only a small fraction of the total mass. Importantly, these scalings have been verified by hundreds of microscopic and phenomenological EOSs available in the literature\,\cite{CLZ23-a,Lat24-talk}, and are also supported by randomly generated meta-model EOSs consistent with all current observational and theoretical constraints\,\cite{CL25-b}, demonstrating their reliability across a wide range of EOS models. Introducing the (dimensionless) log-stability slope\,\cite{CL24-b}
\begin{equation}\label{def-Psi}
\Psi \equiv \frac{2 \varepsilon_{\rm c}}{M_{\rm{NS}}} \frac{\d M_{\rm{NS}}}{\d \varepsilon_{\rm c}}
=2\frac{\d\ln M_{\rm{NS}}}{\d\ln\varepsilon_{\rm{c}}},
\end{equation}
we obtain the expression for the central SSS\,\cite{CL24-a}:
\begin{equation}\label{sc2f}
s_{\rm c}^2 = \x \left( 1 + \frac{1+\Psi}{3} \frac{1+3\x^2+ 4 \x }{1-3\x^2} \right).
\end{equation}

For stable NSs, we have $\Psi \ge 0$; and $\Psi=0$ defines the TOV configuration:
\begin{align}\label{def-TOV-conf}
\text{log-stability slope }\Psi=0&\leftrightarrow\frac{\d M_{\rm{NS}}}{\d\varepsilon_{\rm{c}}}=0\notag\\
&\leftrightarrow\text{ TOV configuration},
\end{align}
which corresponds to the peak position on the NS M-R curve.
Requiring $s_{\rm{c}}^2\leq1$ for TOV NSs then gives the the maximum allowed central EOS-parameter as $\x \lesssim 0.374$. Physically,  this is because the superdense nature of
core NS matter (indicated by the nonlinear dependence of $s_{\rm{c}}^2$ on $\x$) renders the upper bound for $P/\varepsilon$ to be much smaller than unity\,\cite{CLZ23-a},  the basic limit from the principle of causality.  Higher-order corrections in $\hr$ modify the bound on $\x$ only slightly by $\lesssim 3\%$\,\cite{CL25-b}; so we focus on the leading-order expressions in this work.

Notably, the form of the coefficients $b_2$ and $b_4$ is independent of the external input EOS while the coefficient $b_6$ depends on $a_4$; consequently, the ratio $a_2 = b_2/s_{\rm c}^2$ is also independent of the input EOS. This feature enables a robust study of the near-center behavior of $\phi$. In addition, TOV configurations with a vanishing log-stability slope represent a special case that already provides valuable insights into $\phi$ near the NS center, namely $\x=\phi_{\rm c}\lesssim0.374$ as established in the above. Accordingly, in Sections \ref{SEC_3} to \ref{SEC_5}, we focus on $\Psi = 0$, while in Section \ref{SEC_6} we study the effects of a positive log-stability slope $\Psi$ (for non-TOV NSs). With this setup, we are now ready to analyze the behavior of the EOS-parameter $\phi = P/\varepsilon$ in the next section.

\section{Monotonic Decrease of $\phi$ from the NS Center}\label{SEC_3}

\begin{figure}[h!]
\centering
\includegraphics[width=6.2cm]{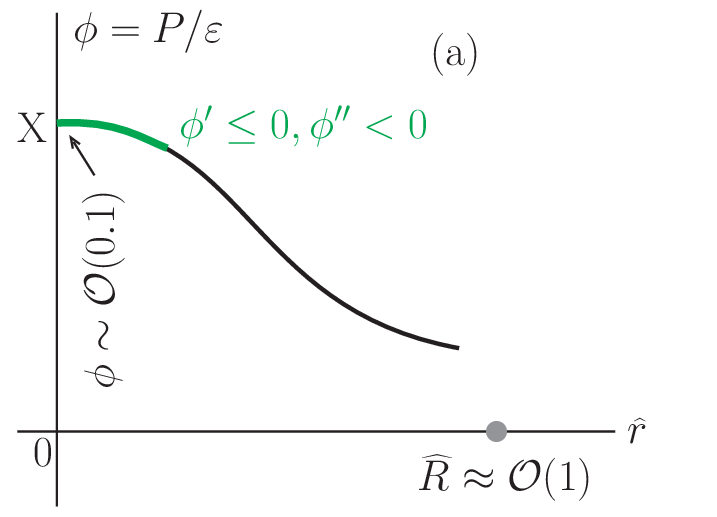}\\[0.5cm]
\includegraphics[width=6.2cm]{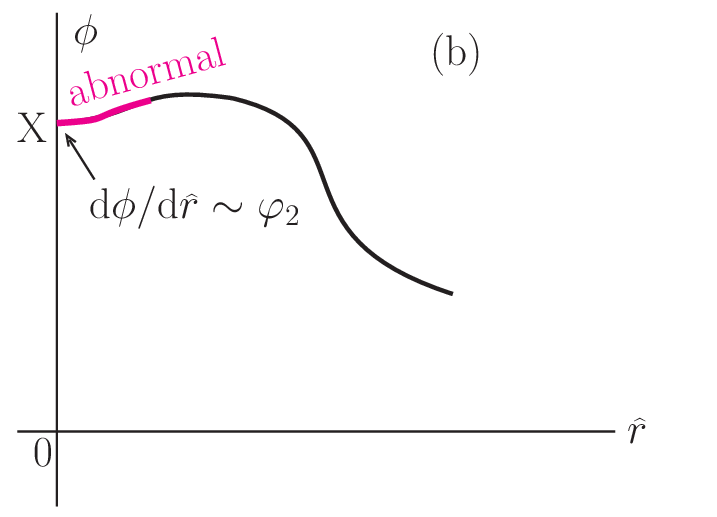}\\[0.5cm]
\includegraphics[width=6.2cm]{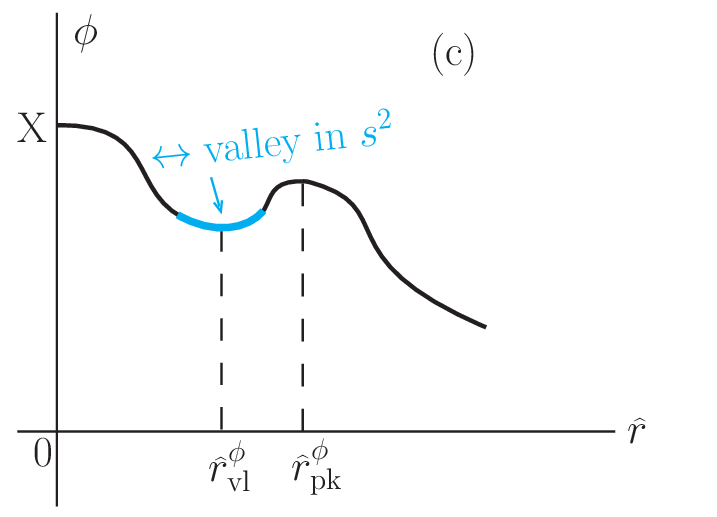}
\caption{(Color Online).  Three possible patterns of the EOS-parameter $\phi$ near NS centers.
Panel (b) is excluded by the expansion of $\phi$ in $\hr$ around the center, while the pattern in panel (c) is incompatible with the presence of a peak in $s^2$ near $\hr=0$ since a valley in $\phi$ will generate a corresponding valley in $s^2$ at a similar energy density, see the analysis of Section \ref{SEC_4}. At the center, $\phi$ approaches $\x$ which is $\lesssim0.374$ to remain casual in NSs at the TOV configuration.
}\label{fig_phi_sk}
\end{figure}

In FIG.\,\ref{fig_phi_sk}, we sketch three typical patterns of the EOS-parameter $\phi = P/\varepsilon = \hP/\heps$ in the vicinity of NS centers. In panel (a), $\phi$ decreases monotonically with the radial coordinate $\hr$ near $\hr \approx 0$. In panel (b), $\phi$ displays an ``abnormal'' outward increase near $\hr \gtrsim 0$. The behavior in panel (c) is similar to that in panel (a) at $\hr \gtrsim0$ but later develops a local valley followed by a local peak (both of which are still assumed to lie within the NS core as indicated by the condition(\ref{RO1})).

\renewcommand*\figurename{\small FIG.}

We discuss panel (a) first, here $\phi'\equiv \d \phi/\d\hr\leq0$.
Since 
\begin{equation}\label{cd-1}
\phi'=\frac{\d\phi}{\d\hr}=\frac{\d\phi}{\d\heps}\frac{\d\heps}{\d\hr},
\end{equation}
and $\heps$ monotonically decreases with $\hr$ (namely $\d\heps/\d\hr\leq0$), the condition $\phi'\leq 0$ is equivalent to $\d\phi/\d\heps=\heps^{-1}(s^2-\phi)\geq0$, or
\begin{equation}\label{cd-2}
s^2\geq\phi\leftrightarrow{s^2}/{\phi}\geq1.
\end{equation}
Using the expressions obtained in the IPAD-TOV approach, we can work out directly that
\begin{align}\label{pk-phi}
\phi(\hr)=&{\hP(\hr)}/{\heps(\hr)}
\approx\x+\varphi_2\hr^2+\varphi_4\hr^4,
\end{align}
with
\begin{align}
\varphi_2=&b_2\left(1-\frac{\x}{s_{\rm{c}}^2}\right),\label{def-varphi2}\\
\varphi_4=&
b_4-a_4\x+\frac{b_2^2\x}{s_{\rm{c}}^4}\left(1-\frac{s_{\rm{c}}^2}{\x}\right).\label{def-varphi4}
\end{align}
Consequently,  $\varphi_2$ is negative definite because $b_2<0$ and $\x/s_{\rm{c}}^2<1$ (see Eq.\,(\ref{sc2f})); and we have perturbatively $\varphi_2\approx-(1+7\x)/24$.
Therefore,
\begin{equation}
\phi''\equiv\frac{\d^2\phi}{\d\hr^2}=
\frac{\d}{\d\hr}\left(2\hr\varphi_2\right)
=
2\varphi_2<0,
\end{equation}
this is also marked in panel (a) of FIG.\,\ref{fig_phi_sk}.

The above demonstrations immediately excludes pattern (b) of FIG.\,\ref{fig_phi_sk}, namely the EOS-parameter $\phi$ is strictly decreasing near the very center of the star; an abnormal behavior, i.e., $\phi$ slightly  raises from $\hr=0$ to some finite $\hr$ is physically inconsistent with the structures of TOV equations.

The next question is whether a possible peak in the EOS-parameter $\phi$ may emerge as $\hr$ goes further toward the surface, as sketched in panel (c) of FIG.\,\ref{fig_phi_sk}. We address this in two steps. First, we show that $\phi$ is likely decreasing near the center (roughly for $\hr\lesssim0.4$ as given in condition (\ref{RO1})) even when higher-order terms such as $\varphi_4 \hr^4$ are included. In the next section, we show that the occurrence of a peak in
$\phi$ at $\hr_{\rm{pk}}^\phi$ (see panel (c) of FIG.\,\ref{fig_phi_sk}) with an unavoidable valley at a smaller $\hr_{\rm{vl}}^\phi<\hr_{\rm{pk}}^{\phi}$,  is fundamentally inconsistent with the pattern in which $s^2$ is peaked around. In other words, if $s^2$ exhibits a peak near the center as recently discussed extensively in the literature\,\cite{ZLi23,Cao23,Mro23,Semp25,Cuce25,Ferr24,Ferr25,Fuji24,Wang24,Ye25}, then $\phi$ is unlikely to display a valley and a following peak at similar positions.

\begin{figure}[h!]
\centering
\includegraphics[width=7.5cm]{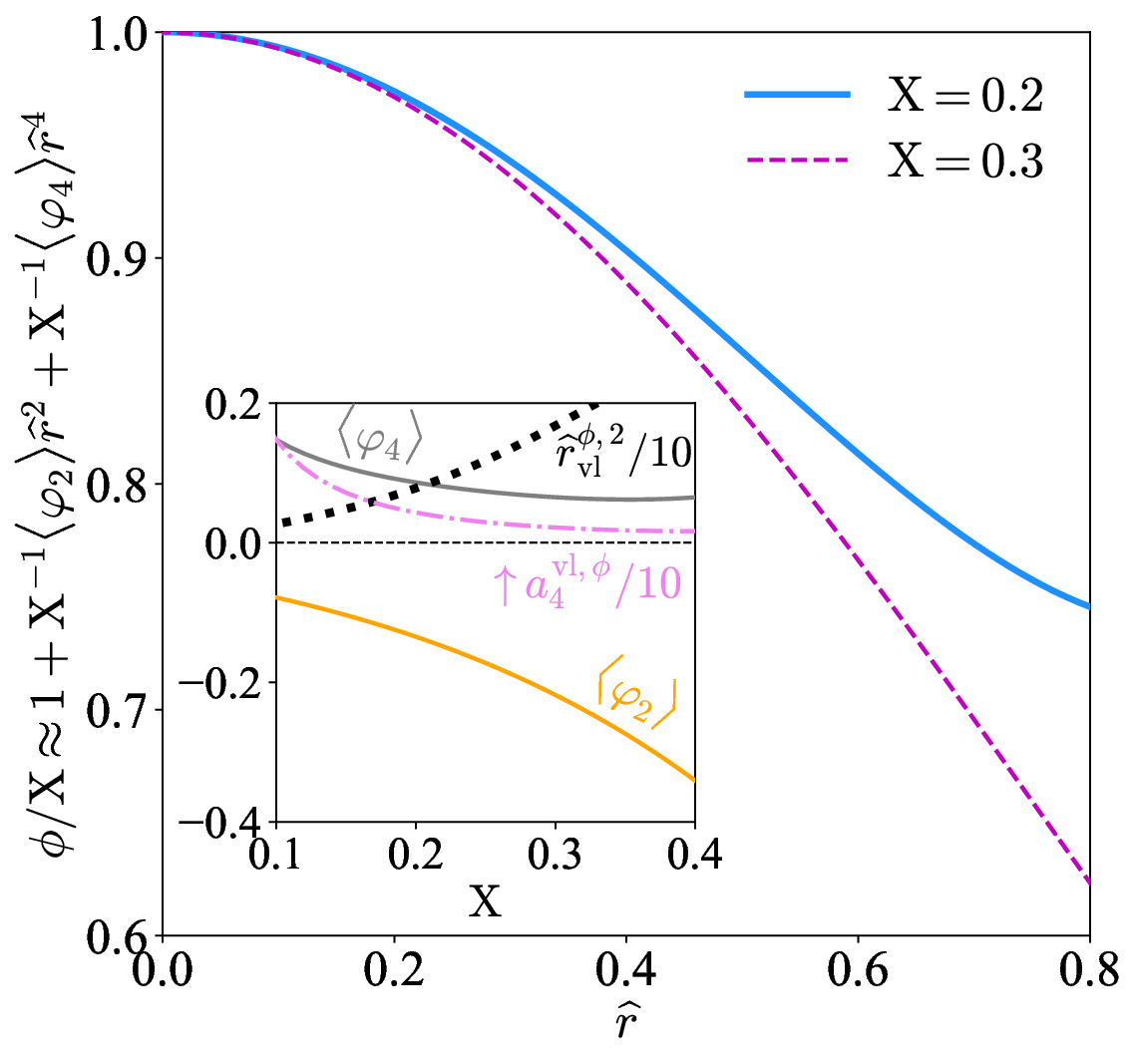}
\caption{(Color Online).   The $\hr$-dependence of $\phi$ to order $\hr^4$ as $\phi\approx \x+\langle \varphi_2\rangle\hr^2+\langle \varphi_4\rangle \hr^4$.
Inset shows the $\x$-dependence of $\langle\varphi_2\rangle$, $\langle\varphi_4\rangle$, $\hr_{\rm{vl}}^{\phi,2}/10$ and $a_4^{\rm{vl},\phi}/10$ (see Eq.\,(\ref{def-a40})).
Although $\langle\varphi_4\rangle$ is positive, the quartic term $\langle\varphi_4\rangle\hr^4$ can hardly change the sign of $\phi-\x$ (near the center) due to the overall smallness of $\hr$.
See the text for details of these quantities.
}\label{fig_phi_aver}
\end{figure}

Since the sign of $\varphi_2$ is deterministic whereas that of $\varphi_4$ depends on the model-dependent coefficient $a_4$ with $|a_4|\sim\mathcal{O}(1)$, we randomly sample $a_4$ within its uncertainty range to examine the averaged behavior of $\varphi_4$. This is equivalent to assuming $\langle a_4\rangle\approx0$, yielding
\begin{equation}
\langle \varphi_2\rangle=b_2\left(1-\frac{\x}{s_{\rm c}^2}\right),\quad
\langle\varphi_4\rangle=b_4+\frac{b_2^2\x}{s_{\rm c}^4}\left(1-\frac{s_{\rm c}^2}{\x}\right).
\end{equation}
The result for $\phi/\x$ up to $\hr^4$ is shown in FIG.\,\ref{fig_phi_aver}. As seen, $\phi$ decreases with $\hr$ near $\hr=0$, despite $\langle\varphi_4\rangle$ being on average positive (inset). The extremum occurs at
\begin{equation}
\hr_{\rm{vl}}^{\phi,2}=-\frac{\langle\varphi_2\rangle}{2\langle\varphi_4\rangle},\quad
\phi_{\rm{vl}}\equiv \phi(\hr^{\phi}_{\rm{vl}})=\x-\frac{\langle\varphi_2\rangle^2}{4\langle\varphi_4\rangle}.
\end{equation}
The inset shows the valley position $\hr_{\rm{vl}}^{\phi,2}/10$ as a function of $\x$; for example, $\hr_{\rm{vl}}^{\phi}\approx1.1$ at $\x\approx0.25$, corresponding to the region very near the NS surface since $\widehat{R}\approx\mathcal{O}(1)$, as shown in Eq.\,(\ref{RO1}). For EOSs with negative $a_4$, the coefficient $\varphi_4$ becomes more positive, and a more pronounced valleyed structure in $\phi$ may develop closer to the center. Taking $a_4\approx-1$ (or $\approx-2$) and $\x\approx0.25$ as an example, we obtain $\hr_{\rm{vl}}^{\phi}\approx0.52$ (or $\approx0.39$), which remains far from the NS center. 
On the other hand, if
\begin{equation}\label{def-a40}
a_4\gtrsim a_4^{\rm{vl},\phi}\equiv \frac{b_4}{\x}+\frac{b_2^2}{s_{\rm{c}}^4}\left(1-\frac{s_{\rm{c}}^2}{\x}\right),
\end{equation}
the coefficient $\varphi_4$ becomes negative; then there would be no valley position $\hr_{\rm{vl}}^{\phi,2}$.
For example, this critical $a_4^{\rm{vl},\phi}$ is about 0.29 if $\x\approx0.25$ is taken, the $\x$-dependence of $a_4^{\rm{vl},\phi}$ is shown by the pink dashed-dotted line in the inset of FIG.\,\ref{fig_phi_aver}.  
These estimates suggest that $\phi$ is unlikely to exhibit a valleyed structure (panel (c) of FIG.\,\ref{fig_phi_sk}) near the NS center. Consequently, a following peaked behavior at $\hr_{\rm{pk}}^{\phi}$ in $\phi$ is also unlikely.

\begin{figure}[h!]
\centering
\includegraphics[width=7.7cm]{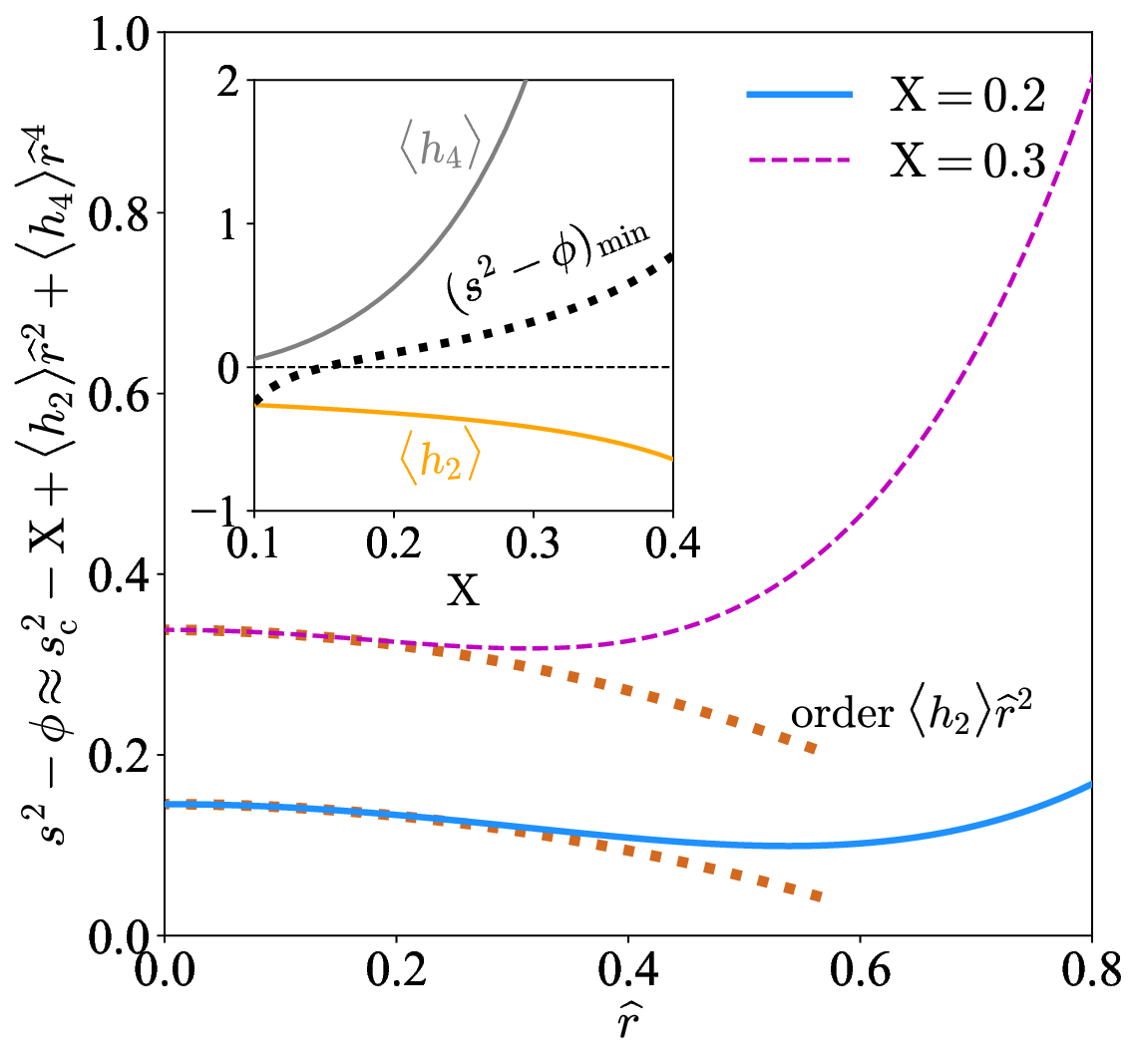}
\caption{(Color Online).  The same as FIG.\,\ref{fig_phi_aver} but for $s^2-\phi$ to order $\hr^4$ as $s^2-\phi\approx s_{\rm{c}}^2-\x+\langle h_2\rangle\hr^2+\langle h_4\rangle \hr^4$ under the assumption that $\langle a_4\rangle\approx\langle a_6\rangle\approx0$.
The inset shows the $\x$-dependence of $\langle h_2\rangle$ and $\langle h_4\rangle$, from which one can find although $\langle h_2\rangle$ is negative the higher order term $\langle h_4\rangle$ is positive.
}\label{fig_s2phi}
\end{figure}

Besides $\phi$ itself, the difference $s^2 - \phi = \heps\d\phi/\d\heps$ captures how the EOS-parameter $\phi$ varies with energy density. It measures the deviation of the local SSS from the mean SSS up to the radial coordinate $\hr$. Condition (\ref{cd-2}) then implies $\phi$ decreases with $\heps$ whenever $s^2 - \phi \ge 0$.
Expanding to $\hr^4$ in the IPAD-TOV framework gives
\begin{align}\label{pk-s2phi}
s^2-\phi \approx s_{\rm c}^2-\x+h_2\hr^2+h_4\hr^4,
\end{align}
where
\begin{align}
h_2 &= \frac{b_2}{s_{\rm c}^2}\!\left[\!\left(\x-s_{\rm c}^2\right)
   +\frac{2s_{\rm c}^4}{b_2^2}\!\left(b_4-a_4s_{\rm c}^2\right)\!\right],\\[2mm]
h_4 &= \frac{b_2^2}{s_{\rm c}^4}\!\Bigg[\!\left(s_{\rm c}^2-\x\right)
   -\frac{s_{\rm c}^4}{b_2^2}\!\left(b_4-a_4\x\right)
   +\frac{3s_{\rm c}^6}{b_2^3}\!\left(b_6-a_6s_{\rm c}^2\right)\notag\\
   &\hspace*{1cm}-\frac{4a_4s_{\rm c}^8}{b_2^4}\!\left(b_4-a_4s_{\rm c}^2\right)\!\Bigg].
\end{align}
The advantage of considering $s^2-\phi$ over $\phi$ is that the former includes even higher-order contributions from the perturbative expansions, namely the $a_6$-term and the $b_6$-term.  Assuming $|a_4|\sim|a_6|\sim\mathcal{O}(1)$ and uniformly distributed in $[-1,1]$, we have $\langle a_6\rangle\approx0$ besides $\langle a_4\rangle\approx$0. Since $h_2$ is linear in $a_4$, the averages are
\begin{align}
\langle h_2\rangle &\approx \frac{b_2}{s_{\rm c}^2}\left(\x-s_{\rm c}^2+\frac{2b_4s_{\rm c}^4}{b_2^2}\right),\\
\langle h_4\rangle &\approx \frac{b_2^2}{s_{\rm c}^4}\left(s_{\rm c}^2-\x-\frac{b_4s_{\rm c}^4}{b_2^2}
   +\frac{3b_6s_{\rm c}^6}{b_2^3}+\frac{4s_{\rm c}^{10}}{3b_2^4}\right),
\end{align}
where $\langle a_4^2\rangle=2^{-1}\int_{-1}^{+1}\!a_4^2\d a_4=1/3$ is used, and the $a_4$ term in $b_6$ is averaged to zero.

FIG.\,\ref{fig_s2phi} shows $s^2-\phi$ up to order $\hr^4$ as a function of $\hr$; the difference $s^2-\phi$ remains positive, indicating that $\phi$ decreases with $\hr$. Although $\langle h_2\rangle<0$, the positive $\langle h_4\rangle$ dominates (inset). The minimum value is
\begin{equation}
{[s^2-\phi]_{\min}\approx s_{\rm c}^2-\x-\frac{\langle h_2\rangle^2}{4\langle h_4\rangle}.}
\end{equation}
As shown in FIG.\,\ref{fig_s2phi}, $s^2-\phi>[s^2-\phi]_{\min}>0$ for $\x\gtrsim0.15$. Since the $\x$ in massive NSs typically exceeds $0.2$\,\cite{CLZ23-a} and the log-stability slope $\Psi$ (of Eq.\,(\ref{def-Psi})) is probably positive for $\x\lesssim0.15$, i.e., the configuration deviates from the TOV point (defined in \,(\ref{def-TOV-conf})), the value of $[s^2-\phi]_{\min}$ would increase correspondingly (as shown by the SSS of Eq.\,(\ref{sc2f}) for $\Psi>0$). 
If nonzero $|a_4|\sim\mathcal{O}(1)$ or $|a_6|\sim\mathcal{O}(1)$ (the form of which is expected to depend on the input EOS) are included, $s^2 - \phi$ remains positive.

\section{A Local Minimum in the EOS-parameter $\phi$ is Incompatible with a Peaked $s^2$ near NS Centers}\label{SEC_4}

In this section, we show in details that a local minimum in the EOS-parameter $\phi$ (as indicated by the position $\hr_{\rm{vl}}$ of panel (c) of FIG.\,\ref{fig_phi_sk}) is basically incompatible with the scenario that a peak exists in $s^2$ near the NS center.
Assume such a local minimum in $\phi$ exists; it then puts a constraint on the coefficient $a_4$ via $\varphi_4>0$:
\begin{empheq}[box=\fbox]{align}
&\phi\approx \x+\varphi_2\hr^2+\varphi_4\hr^4+\cdots,\text{ where }\varphi_2<0\notag\\
\leftrightarrow&\text{a local minimum in $\phi$ near NS centers}\notag\\
\leftrightarrow &\varphi_4=b_4-a_4\x+\frac{b_2^2\x}{s_{\rm{c}}^4}\left(1-\frac{s_{\rm{c}}^2}{\x}\right)>0\notag\\
\leftrightarrow&
a_4<a_4^{\rm{vl},\phi}=\frac{b_4}{\x}+\frac{b_2^2}{s_{\rm{c}}^4}\left(1-\frac{s_{\rm{c}}^2}{\x}\right),\label{cond-1}
\end{empheq}
with $a_4^{\rm{vl},\phi}$ defined in Eq.\,(\ref{def-a40}).
On the other hand,  by considering the SSS perturbatively expanded around $\hr=0$\,\cite{CL24-a}:
\begin{equation}\label{def-l2}
s^2\approx s_{\rm{c}}^2+l_2\hr^2=s_{\rm{c}}^2+\frac{2s_{\rm{c}}^2}{b_2}\left(b_4-a_4s_{\rm{c}}^2\right)\hr^2,
\end{equation}
we can obtain a necessary condition for generating a peaked $s^2$ as\,\cite{CL24-a}:
\begin{empheq}[box=\fbox]{align}
&\text{a peaked structure in $s^2$ near NS centers}\notag\\
\leftrightarrow& l_2>0
\leftrightarrow a_4>a_4^{\rm{pk}}\equiv b_4/s_{\rm{c}}^2>0.\label{cond-2}
\end{empheq}

\begin{figure}[h!]
\centering
\includegraphics[width=7.7cm]{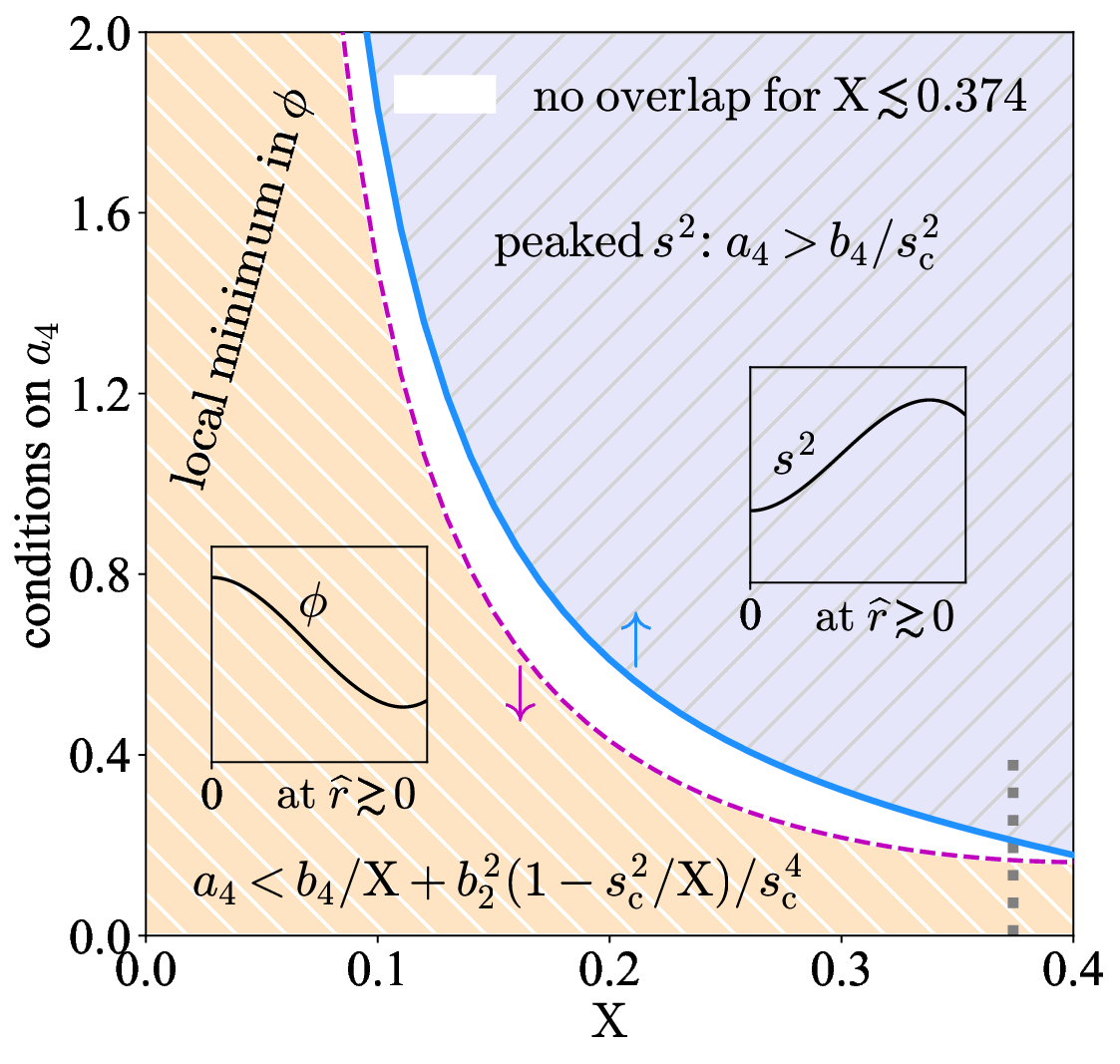}
\caption{(Color Online).  The requirement that there exists a local minimum in the EOS-parameter $\phi$ near NS centers (bisque hatched region) is basically incompatible with the requirement that a peaked $s^2$ exists near NS centers (lavender hatched region); indicated by the fact that there is no overlapped region for $a_4$ parameter from them (white band). The grey dotted line marks the position of $\x\approx0.374$.
Two insets sketch the valley structure of the EOS-parameter $\phi$ and the peaked SSS $s^2$ near the center, respectively.
}\label{fig_a4_phi}
\end{figure}

Therefore, two requirements, namely ``a local minimum existing in the EOS-parameter $\phi$ near NS centers'' of Eq.\,(\ref{cond-1}) and ``a peaked structure in $s^2$ near NS centers'' of Eq.\,(\ref{cond-2}) are basically incompatible with each other:
\begin{equation}
S\cap F=\emptyset, 
\end{equation}
where
\begin{align}
S=&\left\{\left.a_4>a_4^{\rm{pk}}\right|\mbox{peaked $s^2$ near NS centers}\right\},\label{def-setS}\\
F=&\left\{\left.a_4<a_4^{\rm{vl},\phi}\right|\mbox{valleyed $\phi$ near NS centers}\right\},\label{def-setF}
\end{align}
see FIG.\,\ref{fig_a4_phi} for the parameter space for $a_4$ obtained from these two requirements.
It is obvious that there is no overlap between the lavender and bisque regions.
In particular, we can obtain
\begin{align}
&a_4^{\rm{pk}}-a_4^{\rm{vl},\phi}=
\frac{1}{s_{\rm{c}}^2}\left(\frac{s_{\rm{c}}^2}{\x}-1\right)\left(\frac{b_2^2}{s_{\rm{c}}^2}-b_4\right)\notag\\
\approx&\frac{1}{960\x^2}\left(1+17\x+70\x^2+46\x^3-\frac{981}{4}\x^4+\cdots\right),
\end{align}
which is positive definite.

This relationship between the behaviors of $\phi$ and $s^2$, as well as their intrinsic connection, can be understood through the trace-anomaly decomposition of SSS\,\cite{Fuji22}. In this decomposition, we have\,\cite{Fuji22,CL25-a}:
\begin{equation}\label{def-s2-decom}
s^2=\heps\frac{\d\phi}{\d\heps}+\phi=-\heps\frac{\d\Delta}{\d\heps}+\frac{1}{3}-\Delta.
\end{equation}
Physically, Eq.\,(\ref{def-s2-decom}) indicates that $s^2$ depends on both the value of the EOS-parameter $\phi$ and its rate of change with respect to $\heps$. At a local minimum of $\phi$, the slope $\d\phi/\d\heps$ switches from negative to positive, meaning that $\phi$ first decreases and then increases. Because $s^2$ contains the term $\heps\d\phi/\d\heps$, it also transitions from decreasing to increasing as $\heps$ passes through the minimum of $\phi$. Therefore, $s^2$ cannot develop a local maximum (peak) at a similar energy density; instead, it rises through that point, implying that a peak in $s^2$ is impossible near the position where $\phi$ forms a valley.

Actually, one can further show if the EOS-parameter $\phi$ (or the trace anomaly $\Delta$) has a local minimum (or maximum), a valley is expected to appear in the $s^2$ profile. Denoting $\widehat{\varepsilon}_{\phi}^{\rm{vl}}$ as the position of the local minimum in $\phi$ (as illustrated by the valley in panel (c) of FIG.\,\ref{fig_phi_sk}), the position $\widehat{\varepsilon}_{\star}^{\rm{vl}}$ of the corresponding valley in $s^2$ can be obtained as
\begin{equation}\label{fk-1}
\boxed{
\frac{\widehat{\varepsilon}^{\rm{vl}}_{\star}}{\widehat{\varepsilon}^{\rm{vl}}_{\phi}}=\frac{3}{4}\left(1-\frac{1}{k}\right)
+\frac{\sqrt{k^2-2k+9}}{4k},}
\end{equation}
where
\begin{equation}\label{fk-2}
\boxed{
\left.k\equiv\widehat{\varepsilon}^{\rm{vl}}_\phi\frac{\d^3\phi}{\d\widehat{\varepsilon}_{\phi}^{\rm{vl},3}}\right/\frac{\d^2\phi}{\d\widehat{\varepsilon}_{\phi}^{\rm{vl},2}}=\frac{\d}{\d\ln\widehat{\varepsilon}^{\rm{vl}}_\phi}\ln\left(\frac{\d^2\phi}{\d\widehat{\varepsilon}_{\phi}^{\rm{vl},2}}\right).}
\end{equation}
Relations (\ref{fk-1}) and (\ref{fk-2}) are derived by performing a local expansion of the EOS-parameter $\phi$ up to cubic order.
In general, the ratio satisfies
\begin{equation}\label{eps_star}
\boxed{
2^{-1}\lesssim\widehat{\varepsilon}^{\rm{vl}}_{\star}/\widehat{\varepsilon}^{\rm{vl}}_\phi\lesssim1,}
\end{equation}
and
\begin{equation}\label{eps_star2}
\boxed{
\frac{\d^2s^2}{\d\widehat{\varepsilon}_{\star}^{\rm{vl},2}}=\frac{\d^2\phi}{\d\widehat{\varepsilon}_{\phi}^{\rm{vl},2}}\sqrt{k^2-2k+9}\geq2\sqrt{2}\frac{\d^2\phi}{\d\widehat{\varepsilon}_{\phi}^{\rm{vl},2}}.}
\end{equation}
Eqs.\,(\ref{eps_star}) and (\ref{eps_star2}) together indicate that $\heps_{\star}^{\rm{vl}}$ corresponds to a local minimum of $s^2$, located slightly outward (i.e., at a smaller $\heps$) compared to $\heps_{\phi}^{\rm{vl}}$. In other words, the valley in $s^2$ forms somewhat farther from the NS center than the valley in $\phi$. Moreover, the valley in the SSS $s^2$ is sharper than the one in the EOS-parameter $\phi$.

For instance, if $k\approx0$ (i.e., the third-order derivative of $\phi$ is small), one obtains
\begin{equation}
\widehat{\varepsilon}^{\rm{vl}}_{\star}/\widehat{\varepsilon}^{\rm{vl}}_\phi\approx2/3,\quad\text{for }k\approx0.
\end{equation}
Adopting $\widehat{\varepsilon}^{\rm{vl}}_\phi\lesssim\widehat{\varepsilon}_{\rm{c}}=1$ and $\varepsilon_{\rm{c}}/\varepsilon_0\approx6$ for massive NSs, this gives $\widehat{\varepsilon}^{\rm{vl}}_{\star}\approx2\heps^{\rm{vl}}_{\phi}/3\lesssim4\varepsilon_0$, implying that a noticeable valley in $s^2$ appears around four times the nuclear saturation energy density. Moreover, its curvature is about three times larger than that of $\phi$ at $\widehat{\varepsilon}_\phi^{\rm{vl}}$, since Eq.\,(\ref{eps_star2}) yields $\d^2s^2/\d\widehat{\varepsilon}_{\star}^{\rm{vl},2}\approx3\d^2\phi/\d\widehat{\varepsilon}_{\phi}^{\rm{vl},2}$ for $k\approx0$.

Recent investigations of some NS observational constraints indicate that a peaked structure in $s^2$ may emerge near NS centers, although its physical origin and definitive evidence of the exact peak location remain under active discussion\,\cite{ZLi23,Cao23,Mro23,Semp25,Cuce25,Ferr24,Ferr25,Fuji24,Wang24,Ye25}. If this feature is indeed present, the situation represented by case (c) in FIG.\,\ref{fig_phi_sk}, which predicts a valley in $\phi$ near the center, would be essentially ruled out, since a peak or the absence of a valley in $s^2$ is basically incompatible with a valleyed (or peaked) behavior of the EOS-parameter $\phi$ (or the trace anomaly $\Delta$). 

\begin{figure}[h!]
\centering
\includegraphics[width=8.5cm]{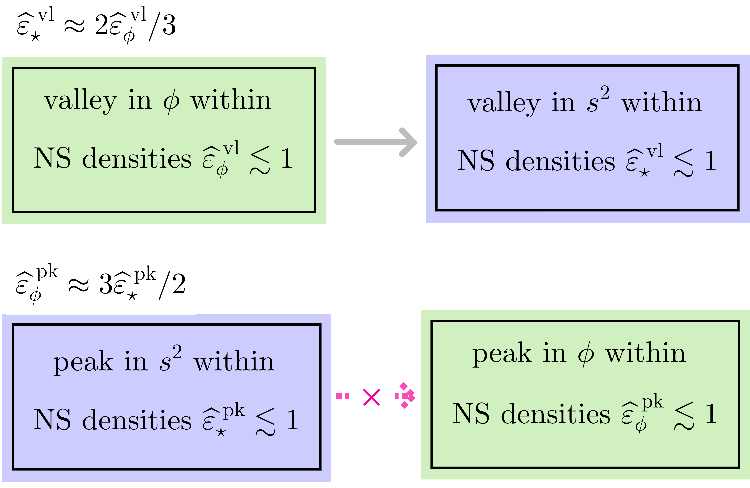}
\caption{(Color Online).  A local minimum (valley) in $\phi$ within NS densities can be used to infer the existence of a valley in $s^2$ at a similar energy density (upper line), because $\heps_{\phi}^{\rm{vl}}\lesssim1$ implies $\heps_{\star}^{\rm{vl}}\lesssim2/3<1$; the inverse, however, does not necessarily hold, since $\heps_{\phi}^{\rm{pk}} \approx3\heps_{\star}^{\rm{pk}}/2$ may exceed unity even if $\heps_{\star}^{\rm{pk}}\lesssim1$.
}\label{fig_vl_pk}
\end{figure}

On the other hand, a peaked $s^2$ near the center can be consistent with a monotonically decreasing EOS-parameter $\phi$ within NS densities: Under the assumption that the third-order derivative of $\phi$ is small, one similarly obtains $\heps_{\star}^{\rm{pk}}/\heps_{\phi}^{\rm{pk}}\approx2/3$, where $\heps_{\star}^{\rm{pk}}$ and $\heps_{\phi}^{\rm{pk}}$ denote the peak positions of $s^2$ and $\phi$, respectively. Consequently, $\heps_{\phi}^{\rm{pk}}\approx3\heps_{\star}^{\rm{pk}}/2$. A peak in $s^2$ within NS densities (i.e., $\heps_{\star}^{\rm{pk}}\lesssim\heps_{\rm{c}}=1$) therefore does not necessarily imply the existence of a corresponding peak in the EOS-parameter $\phi$, since the factor $3/2$ indicates that $\heps_{\phi}^{\rm{pk}}$ may reach or exceed unity. In other words, $\phi$ may attain its maximum either exactly at the NS center or at densities beyond those realized inside NSs, see FIG.\,\ref{fig_vl_pk} for the relation between these positions. In either case, the EOS-parameter $\phi$ decreases monotonically outward.

\section{Neutron Stars at Non-TOV Configurations }\label{SEC_5}

In the previous sections, we examined the behavior of the EOS-parameter $\phi$ near the centers of TOV NSs, where the log-stability slope $\Psi=0$. In this section, we extend the analysis to cases with positive log-stability slope $\Psi$, which correspond to normal NSs below the TOV configuration on the M-R curve, and explore how this affects the corresponding behavior of $\phi$.

Firstly, for non-TOV NSs, the coefficient $\varphi_2$ of Eq.\,(\ref{def-varphi2}) appearing in the EOS-parameter $\phi=P/\varepsilon\approx\x+\varphi_2\hr^2$ near the center can be expressed as
\begin{equation}\label{gt1}
\varphi_2\approx-\frac{1}{24}\frac{1+\Psi}{1+\Psi/4}\left(1+\frac{7+\Psi}{1+\Psi/4}\x\right),
\end{equation}
here the expression for $s_{\rm{c}}^2$ of Eq.\,(\ref{sc2f}) is used.
This shows that the EOS-parameter $\phi$ decreases more rapidly with radial coordinate $\hr$ for non-TOV NSs than for the TOV case (with $\Psi=0$); in particular, we have
\begin{equation}
\varphi_2[\Psi]-\varphi_2[\Psi=0]\approx-\frac{1}{32}\frac{\Psi}{1+\Psi/4}<0.
\end{equation}

Secondly, we generalize the expression for the possible valley position of $\phi$, as illustrated in FIG.\,\ref{fig_phi_aver}:
\begin{equation}\label{gt2}
\hr_{\rm{vl}}^{\phi,2}=-\frac{\langle\varphi_2\rangle}{2\langle\varphi_4\rangle}\approx\frac{20}{11}
\frac{(1+\Psi)(1+\Psi/4)}{1-\Psi/11}\x.
\end{equation}
A positive log-stability slope $\Psi$ shifts the valley position (if it exists) outward toward the NS surface, indicating that $\phi$ is more likely to decrease monotonically with $\hr$ when $\Psi>0$.
The critical value $a_4^{\rm{vl},\phi}$ of Eq.\,(\ref{def-a40}) beyond which the valley structure in  the EOS-parameter $\phi$ disappears similarly becomes
\begin{equation}\label{gt3}
a_4^{\rm{vl},\phi}\approx\frac{11}{960}\frac{1-\Psi/11}{(1+\Psi/4)^2}\frac{1}{\x^2},
\end{equation}
which is smaller than in the $\Psi=0$ case. Hence, the parameter space without a valley in $\phi$ becomes larger, making a valley structure even less likely.

Thirdly, we evaluate the minimum of $s^2-\phi$ using expressions for $\phi$ and $s^2$ up to order $\hr^4$, which determines whether $\phi$ develops a peak or valley near the center (see FIG.\,\ref{fig_s2phi}).
For positive log-stability slope $\Psi$:
\begin{equation}\label{gt4}
[s^2-\phi]_{\min}=s_{\rm{c}}^2-\x-\frac{\langle h_2\rangle^2}{4\langle h_4\rangle}\approx\frac{311+178\Psi+11\Psi^2}{165-15\Psi}\x,
\end{equation}
which is more likely to be positive than in the $\Psi=0$ case (see the inset of FIG.\,\ref{fig_s2phi}).
A positive $s^2(\hr)-\phi(\hr)$ implies that $\phi(\hr)$ decreases from $\hr$ correspondingly.

\begin{figure}[h!]
\centering
\includegraphics[width=7.7cm]{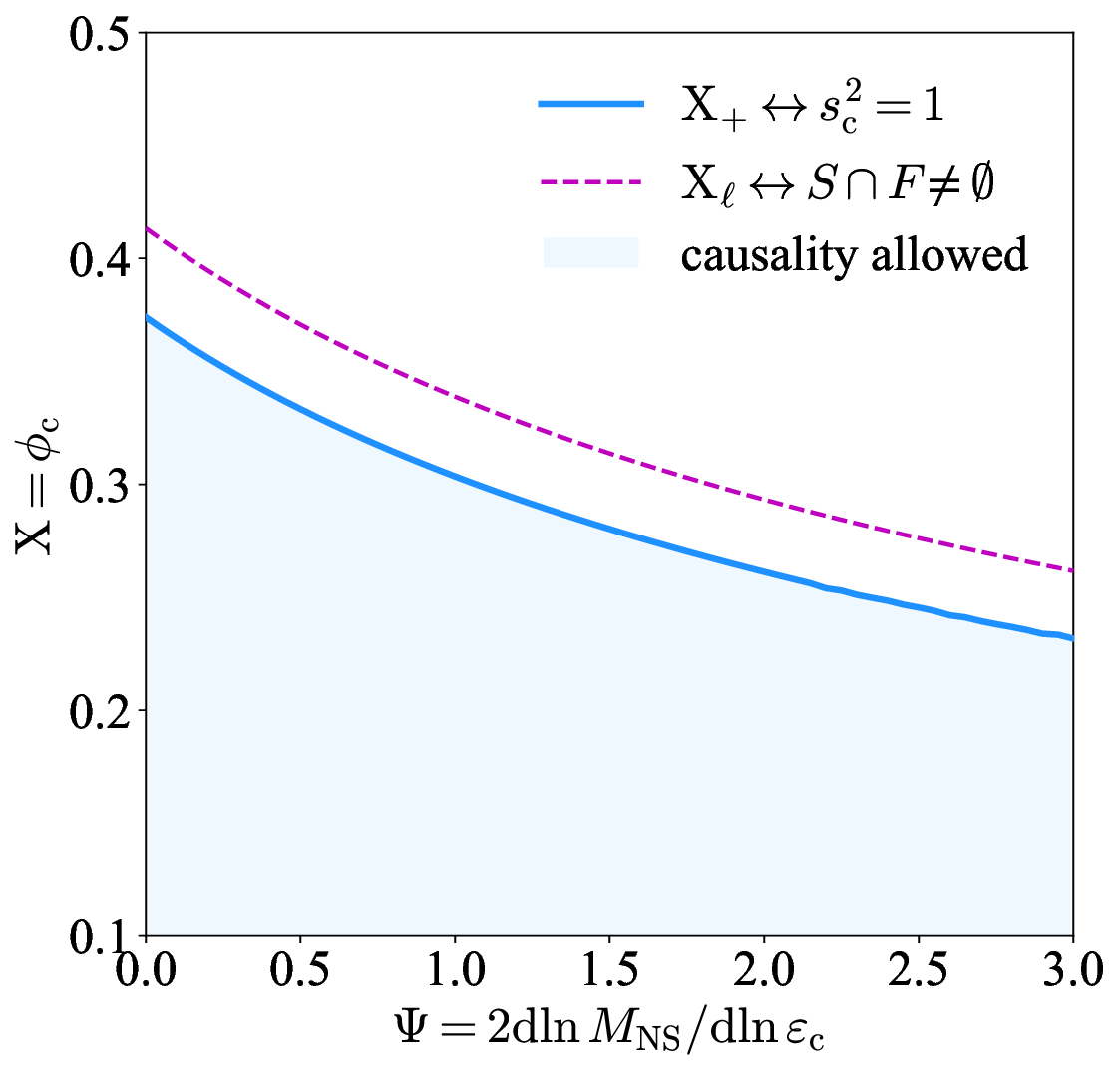}
\caption{(Color Online).  Dependence of $\x_+$ and $\x_{\ell}$ on the log-stability slope $\Psi$; here $\x_{\ell}$ is defined as the smallest value of $\x$ above which the two parameter sets $S$ of (\ref{def-setS}) and $F$ of (\ref{def-setF}) start to intersect.
}\label{fig_UpX}
\end{figure}

\begin{figure}[h!]
\centering
\includegraphics[width=3.8cm]{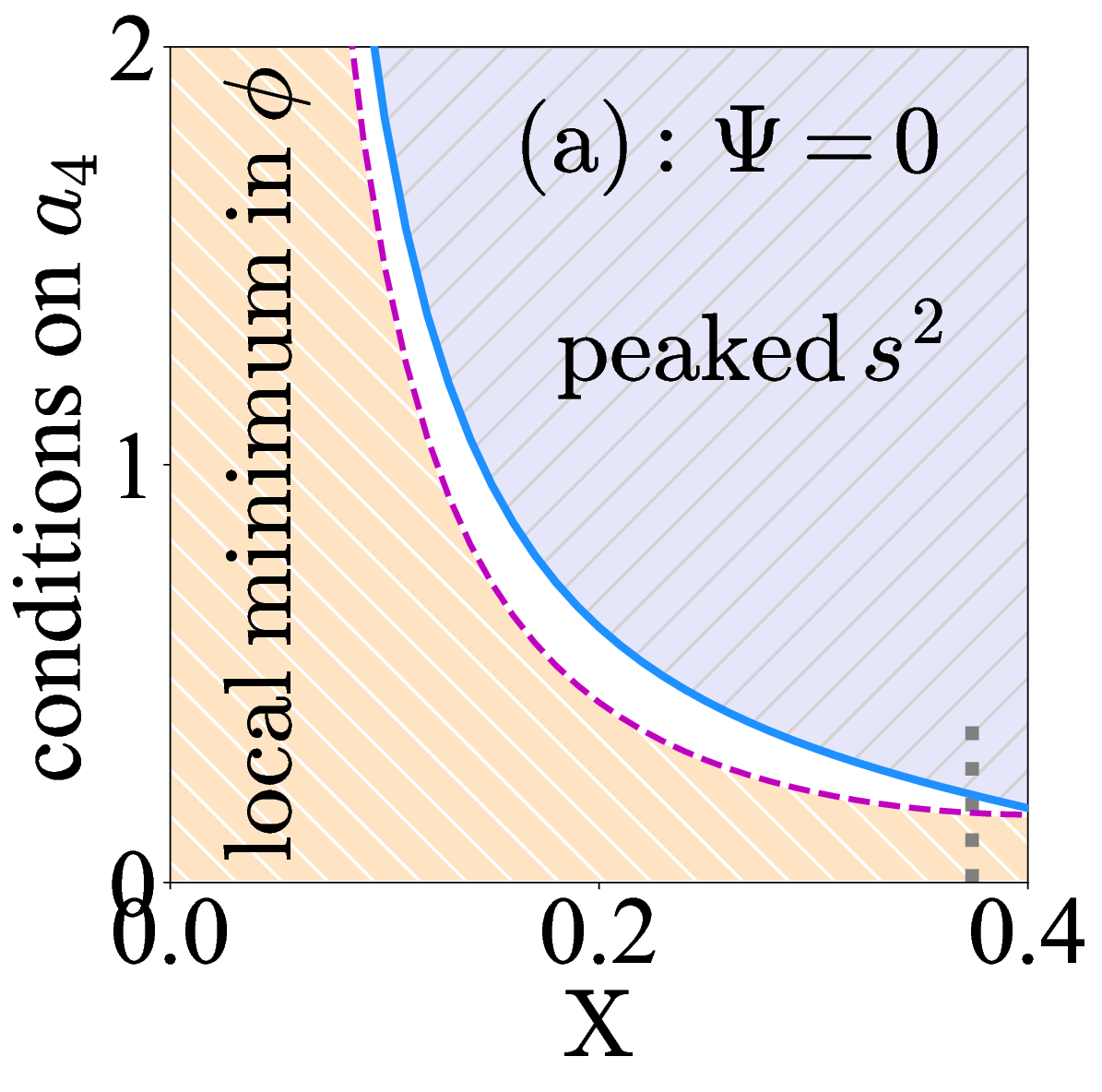}\quad
\includegraphics[width=3.8cm]{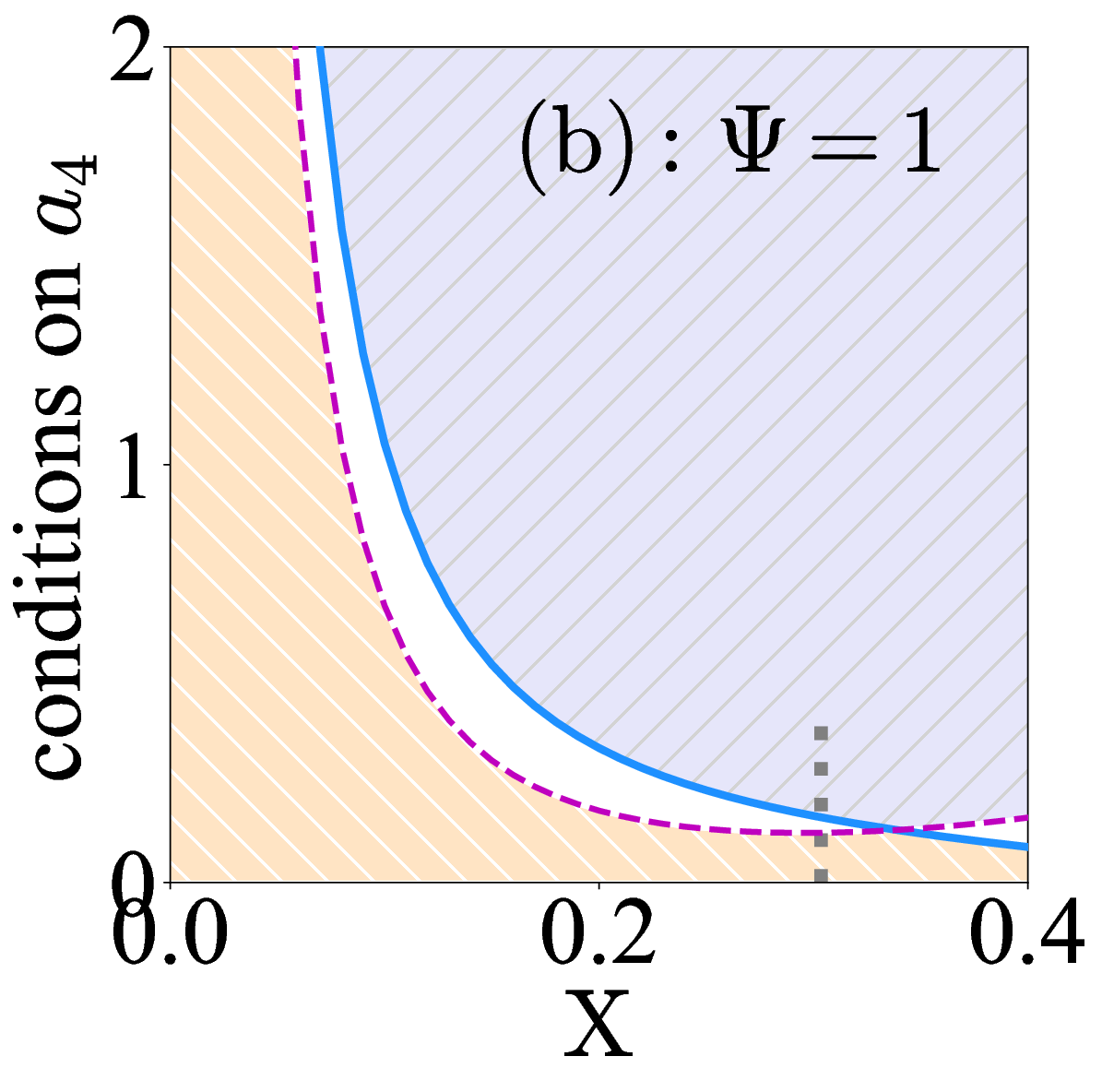}\\
\includegraphics[width=3.8cm]{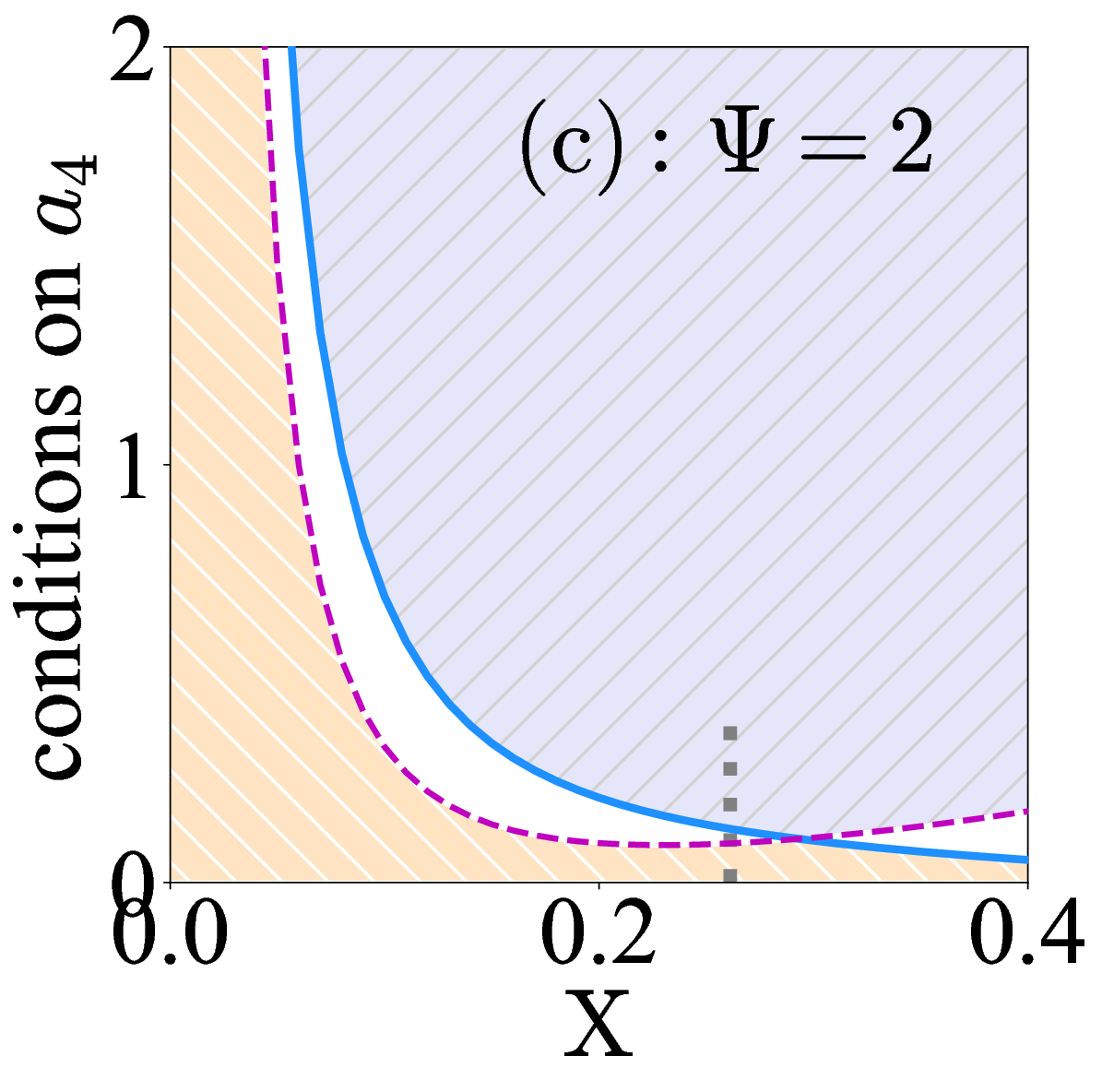}\quad
\includegraphics[width=3.8cm]{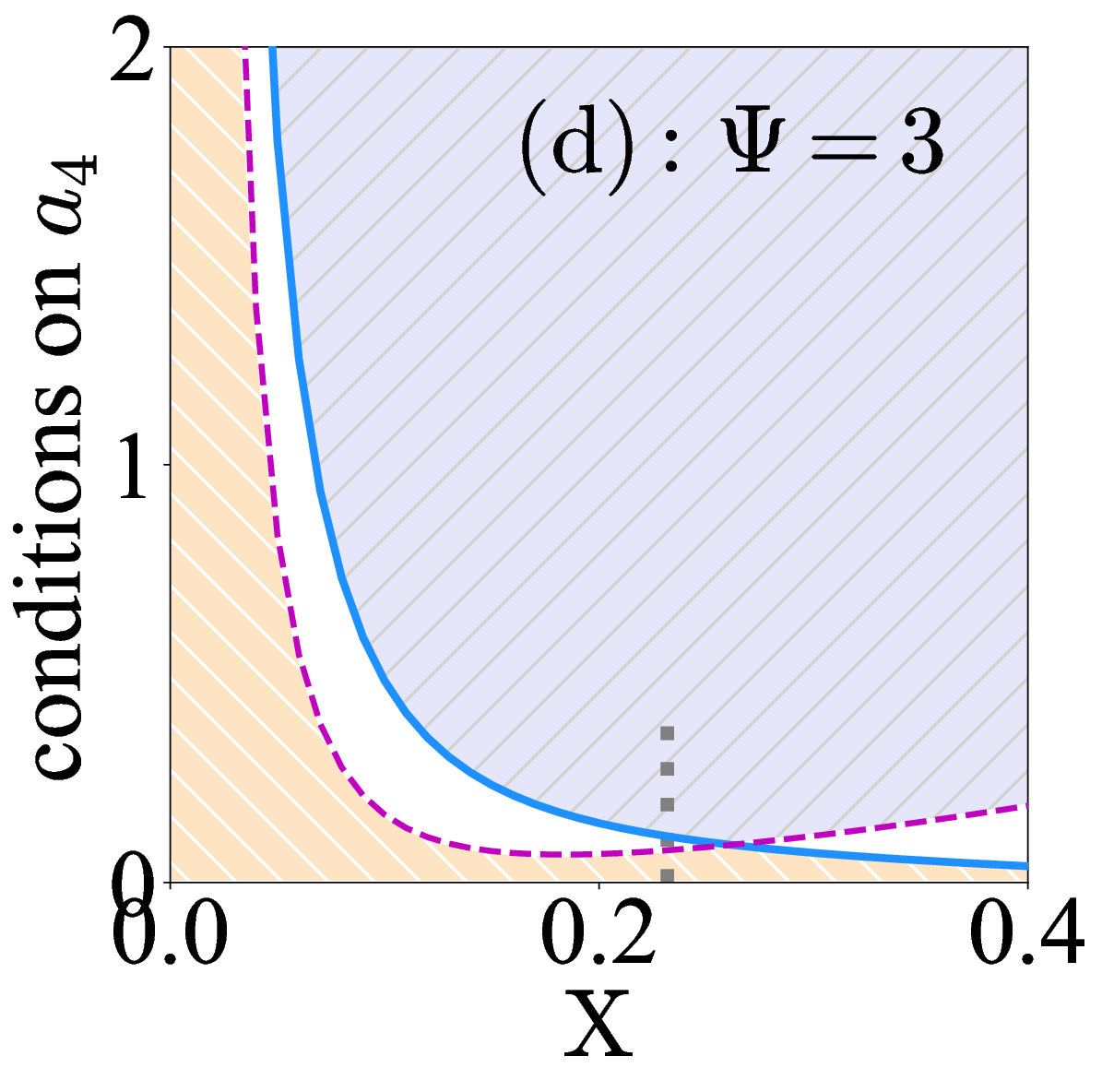}
\caption{(Color Online).  Same as FIG.\,\ref{fig_a4_phi}, but for different values of $\Psi$. Below the causality boundary $\x_+$, the two sets $S$ of (\ref{def-setS}) and $F$ of (\ref{def-setF}) do not intersect.
The grey vertical lines mark the $\x_+$ for each $\Psi$.
}\label{fig_a4_phi_abcd}
\end{figure}

Eqs.\,(\ref{gt1})-(\ref{gt4}) are perturbative in $\x$. Numerical evaluations confirm the same qualitative trend: a positive $\Psi$ enhances the tendency for the EOS-parameter $\phi$ to decrease monotonically with radial coordinate. Therefore, for relatively light NSs away from the TOV configuration, such as the canonical ones, the $\phi$ is expected to be a monotonically decreasing function over a wide radial range, consistent with existing studies\,\cite{Ecker23}.

We notice that although $\phi$ is likely decreasing from the NS center and takes its maximum at $\hr=0$ according to the IPAD-TOV approach, we cannot rule out the possibility of ripples in $\phi$ near the surface, which may be largely influenced by the crust EOS and lie beyond the predictive power of the IPAD-TOV method.

Finally, we confirm that introducing a positive log-stability slope $\Psi$ does not alter the conclusion of FIG.\,\ref{fig_a4_phi}. In that figure, when $\x$ becomes sufficiently large, the parameter regions that allow a peaked $s^2$ and those that allow a valleyed $\phi$ begin to overlap. For TOV NSs ($\Psi=0$), this overlap first appears at $\x_{\ell}\approx0.413$, which is safely above the causality limit $\x_+\approx0.374$ (defined by $s_{\rm{c}}^2=1$).
When a positive log-stability slope $\Psi$ is introduced, both $\x_{\ell}$ and the causality boundary $\x_+$ are modified to smaller values, while the inequality $\x_{\ell}>\x_+$ always remains valid, as shown in FIG.\,\ref{fig_UpX}. This indicates that, even with a positive log-stability slope $\Psi$, the parameter space for $a_4$ where $s^2$ becomes peaked never coincides with the parameter space where the EOS-parameter $\phi$ develops a local minimum, meaning that these two features cannot occur simultaneously within the causal domain.
This conclusion is essentially independent of $\Psi$, since it follows from the trace-anomaly decomposition of $s^2$ showed above. The parameter spaces for $a_4$ associated with the sets $S$ of (\ref{def-setS}) and $F$ of (\ref{def-setF}) for several representative log-stability slope $\Psi$ values are shown in FIG.\,\ref{fig_a4_phi_abcd}.

\section{Potential Effects of pQCD Constraints on the Near-center Behavior of the EOS-parameter $\phi$}\label{SEC_6}

At ultra-high densities, pQCD provides a robust and effective constraint on quark matter EOS\,\cite{Bjorken83,Kurkela10,Gorda21PRL,Gorda23PRL,Gorda23,Komo22}. 
In this regime, the strong coupling constant becomes small, allowing the pressure to be computed reliably from first principles\,\cite{Bjorken83,Kurkela10}. 
Despite the significance of the pQCD method, there is currently no consensus on the relevance of pQCD for NSs, since the densities in typical NS cores (roughly $4\lesssim\varepsilon_{\rm c}/\varepsilon_0\lesssim8$) may not be high enough for perturbative methods to be fully reliable\,\cite{Som23-a,Zhou25}.
In addition, pQCD sets an upper bound (1/3) on the EOS-parameter $\phi$ at extremely high densities $\gtrsim50\varepsilon_0$\,\cite{Bjorken83}. This limits asymptotically how stiff the EOS can be. Nevertheless, evaluating its impact on NSs requires a model-dependent interpolation between the low- and intermediate-density nuclear physics and the ultra-high density pQCD.

\begin{figure}[h!]
\centering
\includegraphics[width=7.3cm]{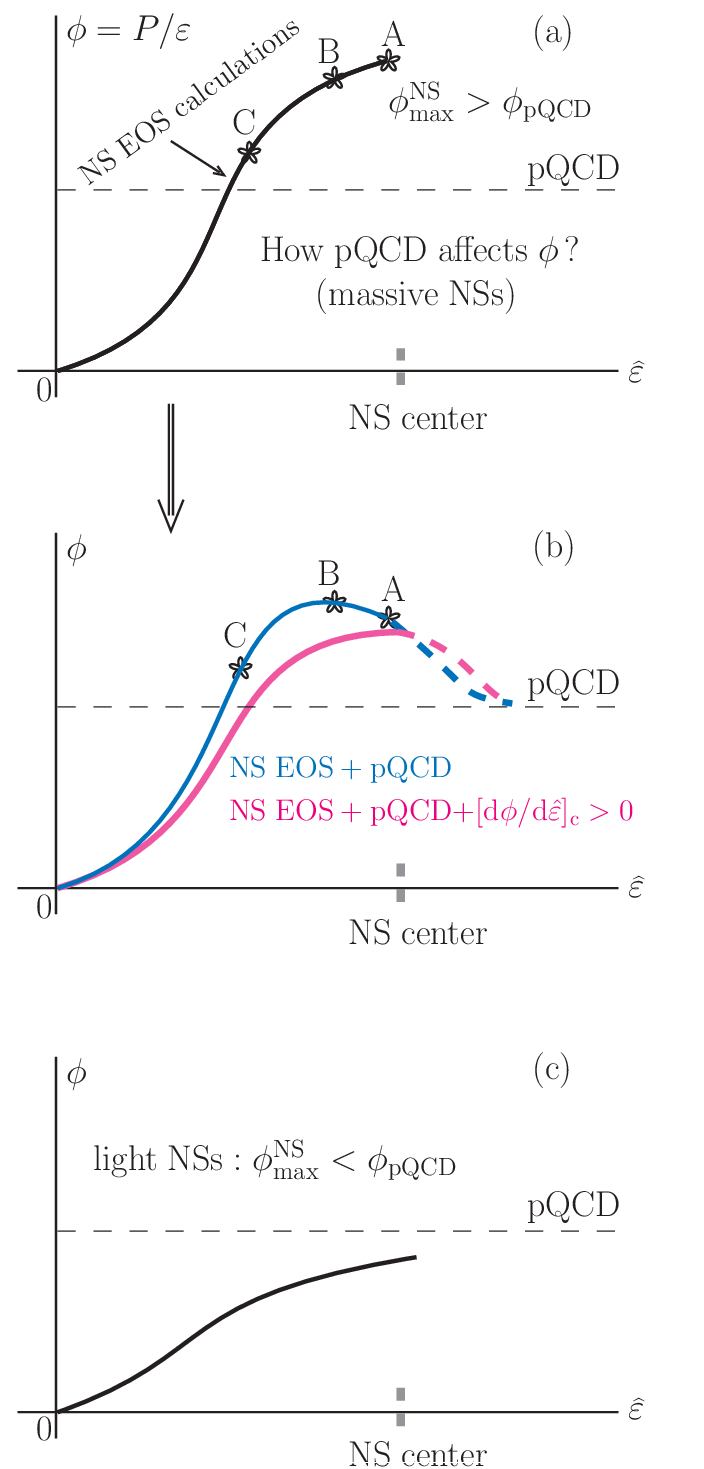}
\caption{(Color Online).  The pQCD effects at extremely high densities modify the EOS-parameter $\phi$ more strongly at point ``A'' than at point ``B'', since ``A'' is closer to the NS center (see panel (a)). For light NSs,  the maximum $\phi_{\max}^{\rm{NS}}$ may remain well below the pQCD limit ($\phi_{\rm{pQCD}}$), in which case the pQCD constraint has minimal impact on the $\phi$ profile (see panel (c)). When analyzing the EOS-parameter $\phi$ near NS centers, in addition to the NS EOS and pQCD constraints, the extra condition $\left[ \d\phi / \d\widehat{\varepsilon} \right]_{\rm c} > 0$ should also be imposed.
}\label{fig_phi_pQCD}
\end{figure}

When pQCD constraints at asymptotically high densities are included, they can modify the EOS and therefore the behavior of the EOS-parameter $\phi$ in NSs\,\cite{Semp25}. For example, the value of $\phi$ at point ``A'' (closer to the center than point ``B'') is affected more strongly than at ``B'', which can produce an outward-pointing bump in the $\phi(\hr)$ profile, as sketched in panel (b) of FIG.\,\ref{fig_phi_pQCD}. By contrast, the EOS-parameter $\phi$ at point ``C'' (close to NS surface) is almost unaffected by the pQCD constraints. Thus a possible peak in $\phi$ is likely a combined effect of the NS EOS and the pQCD limit; without the latter, the $\phi$ profile would more likely resemble panel (a).

The mechanism producing a possible peak in the EOS-parameter $\phi$ differs from that for a peak in $s^2$. The curvature controlling coefficient, $l_2 = (2 s_{\rm c}^2 / b_2)(b_4 - a_4 s_{\rm c}^2)$ of $s^2(\hr)\approx s_{\rm c}^2+l_2\hr^2+\cdots$, can itself be positive, for example, due to the strong-field general relativity\,\cite{CL24-a} or nuclear matter EOS effects\,\cite{ZLi23,Wang24}, so $s^2$ may develop a peak intrinsically. In contrast, $\varphi_2 = b_2 (1 - \x / s_{\rm c}^2)$ of $\phi(\hr)\approx\x+\varphi_2\hr^2+\cdots$ is definitely negative, so an intrinsic peak in $\phi$ near the center cannot arise from the local expansion alone; the pQCD constraint is probably responsible for the extruded peak in $\phi$.
For light NSs, the maximum $\phi_{\max}^{\rm{NS}}$ may remain well below the pQCD bound, in which case enforcing pQCD does not induce a peak; this situation is illustrated in panel (c) of FIG.\,\ref{fig_phi_pQCD}. It is worth emphasizing that a peaked structure in $\phi$ is closely tied to a negative trace anomaly, $\Delta = 1/3 - \phi$. As already discussed, at sufficiently high densities one generally has $\Delta \to 0$, a behavior that inevitably generates a local minimum in $\Delta$. For instance, Ref.\,\cite{Fuji22} provides an insightful analysis of this issue: by scanning the EOS-parameter space through TOV solutions, it found that $\Delta$ may become negative and demonstrated that certain maximally stiff EOSs can indeed yield negative values of $\Delta$. In this sense, a negative $\Delta$ may serve as an underlying origin of a peaked feature in $\phi$ at some densities inside NSs.

We summarize below the possible near-center patterns of the EOS-parameter $\phi(\hr)$ for massive NSs:
\begin{enumerate}[label=(\alph*),leftmargin=*]
\item Near the center, $\phi(\hr) \approx \x + \varphi_2\hr^2$ with $\varphi_2 < 0$. Therefore $\phi$ must decrease with $\hr$ immediately outward from $\hr=0$. An outward-increasing profile at the origin is excluded (see panel (b) of FIG.\,\ref{fig_phi_sk}).

\item The existence of a local minimum of $\phi(\hr)$ at $\hr \neq 0$ would imply a corresponding valley in $s^2$ at a similar energy density. The presence of a peaked $s^2$ near the center and the absence of such valleys in existing predictions using EOSs consistent with NS observations make this scenario unlikely (see panel (c) of FIG.\,\ref{fig_phi_sk}).
\end{enumerate}

Combining these arguments, $\phi(\hr)$ is expected to be monotonically decreasing near the center  (panel (a) of FIG.\,\ref{fig_phi_sk}); consequently $\Delta \gtrsim \Delta_{\rm c} = 1/3 - \x$\,\cite{CL24-b}. If $\x$ in massive NSs exceeds the pQCD limit, the pQCD constraint can effectively modify the inferred $\phi(\hr)$ and induce a peak. We emphasize that such a peak is the combined consequence of (i) interpolating NS EOSs forcefully to meet the pQCD constraint at asymptotically high densities and (ii) a large central EOS-parameter $\x$ relevant only for massive NSs.
Moreover, one should ensure the combined constraints:
\begin{equation}
\boxed{
\text{NS EOS calculations} + \text{pQCD} + \left.\frac{\d\phi}{\d\widehat{\varepsilon}}\right|_{\rm c} > 0,}
\end{equation}
when studying the near-center behavior of the EOS-parameter $\phi$.

\section{Mean Stiffness of NS Core Matter}\label{SEC_7}

As discussed above, the EOS-parameter $\phi$ is a measure of the mean stiffness of NS matter from its surface up to an inner position $\hr$\,\cite{Marc24,Saes24}. Complementary to $\phi$, we can define the mean stiffness $\Phi$ of NS core matter by integrating the SSS $s^2$ starting from its center to the position $\heps(\hr)$:
\begin{equation}\label{def-CMS}
\Phi=\Phi(\heps)=
\frac{1}{\heps-1}\int^{\heps}_1\frac{\d \hP}{\d\heps'}\d\heps'=\frac{\x-\hP}{1-\heps},
\end{equation}
here $1-\heps=\heps_{\rm{c}}-\heps>0$.
The two mean stiffnesses, $\Phi$ (inner) and $\phi$ (outer), satisfy the following sum rule:
\begin{equation}\label{def-SR}
    \Phi(1-\heps)+\phi\heps=\x.
\end{equation}
Then for a given NS (its central EOS-parameter $\x$ at $r=0$ is fixed), $\Phi$ and $\phi$ behave oppositely as $\heps$ varies from the center ($\heps=\heps_{\rm{c}}=1$) to the surface ($\heps\to0$), reflecting the trade-off between the two contributions to the central EOS-parameter $\x$.

According to the definition (\ref{def-CMS}) for the complementary mean stiffness $\Phi$ of NS cores we have
\begin{equation}\label{def-Phi1}
\frac{\d\Phi}{\d\heps}=\frac{\Phi-s^2}{1-\heps}\approx-\frac{1}{2}\frac{l_2\hr^2}{1-\heps},
\end{equation}
where the coefficient $l_2$ is defined in Eq.\,(\ref{def-l2}), and the second relation holds near $\hr\approx0$. The sign of $\d\Phi/\d\heps$ depends on $l_2$: if $l_2>0$ (corresponding to a peak in $s^2$ near the center), $\d\Phi/\d\heps<0$, and the inner mean stiffness $\Phi$ does not take its maximum at the center; if otherwise $l_2<0$, $\Phi$ is maximized at the center.

Physically, this behavior reflects how the complementary mean stiffness $\Phi$ integrates the local stiffness $s^2$ from the center outward. If $s^2$ exhibits a local peak near the center ($l_2>0$), the contribution from the immediate surroundings of the center is higher than the average over the remaining inner region, causing $\Phi$ to initially increase with decreasing $\heps$ (i.e., moving outward); in this case, the maximum of $\Phi$ occurs away from the center. Conversely, if $s^2$ decreases outward ($l_2<0$), the stiffness near the center dominates the integral, and $\Phi$ naturally reaches its maximum at $\heps_{\rm{c}}=1$. Therefore, the shape of $s^2$ near the center directly determines whether the $\Phi$ is centrally peaked or shifted outward.

In contrast, the outer mean stiffness $\phi = P/\varepsilon = \hP/\heps$\,\cite{Marc24,Saes24} accumulates $s^2$ from the stellar surface inward. Near the center, $\d\phi/\d\heps \approx \heps^{-1}(s_{\rm c}^2 - \x)$ is positive because the central stiffness $s_{\rm c}^2$ is larger than the overall mean stiffness or center EOS-parameter $\x$. As a result, $\phi$ monotonically decreases outward from the center. The key difference is that the EOS-parameter $\phi$ measures the stiffness already integrated from the surface, while $\Phi$ measures the remaining average from the center outward, as indicated by the sum rule (\ref{def-SR}). Consequently, local peaks or valleys in $s^2$ near the center can shift the maximum of $\Phi$ away from the center, whereas $\phi$ always decreases outward from $\hr \approx 0$, independent of local variations in $s^2$.
In other words, even if structures such as peaks or valleys may appear in $s^2$ in the outer layers (e.g., $2\widehat{R}/3 \lesssim \hr \lesssim \widehat{R}$), the integration of $s^2$ over $\heps$ effectively smooths out these surface features. The behavior of $\phi$ near the center is mainly governed by the intrinsic structure of the TOV equations rather than by local fluctuations in $s^2$ far away from the center.

\section{Conclusions}\label{SEC_8}
In conclusion, by analyzing the near-center behavior of the EOS-parameter $\phi = P/\varepsilon$ and the corresponding trace anomaly $\Delta = 1/3 - \phi$ in NS cores using the IPAD-TOV approach, we found that the $\phi$ decreases monotonically outward from the NS center, corresponding to a monotonic increase of $\Delta$.  A local minimum in $\phi$ is basically incompatible with a peaked SSS $s^2$ at similar energy densities. Observational evidence of a peaked $s^2$ profile near NS centers therefore excludes a valley and consequently a following peak in the EOS-parameter $\phi$. Enforcing pQCD constraints at asymptotically high densities can modify the EOS-parameter $\phi$ in massive NSs and generate an outward bump, but the overall decreasing trend of $\phi$ should remain preserved. This trend is essential for understanding the microscopic properties and stiffness of superdense matter in NS cores. Our results offer input EOS-independent new insights linking nuclear physics, NS astrophysics, and pQCD through the EOS-parameter $\phi$.

\section*{Acknowledgement} We would like to thank Xavier Grundler, Wen-Jie Xie, Nai-Bo Zhang and Zhen Zhang for helpful discussions. This work was supported in part by the National Natural Science Foundation
of China under contract No. 12147101, the U.S. Department of Energy, Office of Science, under Award Number DE-SC0013702, the CUSTIPEN (China-U.S. Theory Institute for Physics with
Exotic Nuclei) under the US Department of Energy Grant No. DE-SC0009971.

\section*{DATA AVAILABILITY}
The data that support the findings of this article will be openly available\,\cite{data}.

\end{document}